\documentclass[lettersize,journal]{IEEEtran}
\usepackage{amsmath,amsfonts}
\usepackage{algorithm}
\usepackage{algpseudocode}
\usepackage{array}
\usepackage[caption=false,font=normalsize,labelfont=sf,textfont=sf]{subfig}
\usepackage{textcomp}
\usepackage{stfloats}
\usepackage{url}
\usepackage{verbatim}
\usepackage{graphicx}
\usepackage{xcolor}
\usepackage{orcidlink}

\def\BibTeX{{\rm B\kern-.05em{\sc i\kern-.025em b}\kern-.08em
    T\kern-.1667em\lower.7ex\hbox{E}\kern-.125emX}}
\usepackage{balance}
\begin{document}
\title{Self-organizing railway traffic management}
\author{Federico Naldini, Fabio Oddi, Leo D'Amato, Gr\'egory Marli\`ere, Vito Trianni, Paola Pellegrini
\thanks{Federico Naldini\,\orcidlink{0000-0002-2144-9937}, Gr\'egory Marli\`ere\,\orcidlink{0009-0005-9454-3835} and Paola Pellegrini\,\orcidlink{0000-0002-6087-651X} are with Univ Gustave Eiffel, COSYS-ESTAS, F-59650 Villeneuve d’Ascq, France.}
\thanks{Fabio Oddi\,\orcidlink{0009-0001-8379-0209}, Leo D'Amato\,\orcidlink{0009-0007-4531-5486}, Vito Trianni\,\orcidlink{0000-0002-9114-8486} are with Institute for Cognitive Sciences and Technologies, CNR, Via Giandomenico Romagnosi 18A, 00196, Rome, Italy.} 
\thanks{Leo D'Amato is with Dipartimento di Automatica e Informatica, Politecnico di Torino, Corso Duca degli Abruzzi, 24, 10129 Torino, Italy.}
\thanks{Fabio Oddi is with Dipartimento di Ingegneria Informatica, Automatica e Gestionale, La Sapienza Università di Roma, Via Ariosto 25, 00185, Rome, Italy.}
}

%
%

\markboth{PREPRINT --- IEEE Transactions on Intelligent Transportation Systems}%
{Naldini et al.: Self-organizing railway traffic management }

\maketitle

\begin{abstract} 
Improving traffic management in case of perturbation is one of the main challenges in today's railway research. The great majority of the existing literature proposes approaches to make centralized decisions to minimize delay propagation. In this paper, we propose a new paradigm to the same aim: we design and implement a modular process to allow trains to self-organize. This process consists in having trains identifying their neighbors, formulating traffic management hypotheses, checking their compatibility and selecting the best ones through a consensus mechanism. Finally, these hypotheses are merged into a directly applicable traffic plan. In a thorough experimental analysis on a portion of the Italian network, we compare the results of self-organization with those of a state-of-the-art centralized approach. In particular, we make this comparison mimicking a realistic deployment thanks to a closed-loop framework including a microscopic railway simulator. The results indicate that self-organization achieves better results than the centralized algorithm, specifically thanks to the definition and exploitation of the instance decomposition allowed by the proposed approach.
\end{abstract}


\section{Introduction}\label{sec:intro}
\IEEEPARstart{R}{ailway} traffic is often affected by unexpected events. These events can be perturbations or disturbances, i.e., minor deviations from the timetable, or disruptions, i.e., major degrading events~\cite{CacchianiSurvey2014, Jenelius2015_MATTSSON201516}. While disruptions impose significant changes to the timetable to be agreed upon by different stakeholders, perturbations  require train routes and schedules to be carefully selected to limit delay propagation. This selection is currently made by dispatchers, with very limited decision support tools. 
To deal with perturbations,  from the traffic control centers, dispatchers make decisions on traffic evolution in a control area based on experience and a set of simple heuristic rules. The railway community, including public authorities, industry and academia, is today very actively aiming at the introduction of intelligent decision support tools in traffic control centers. Among other initiatives, the European Joint Undertakings Shift2Rail \cite{S2R} and EuropeRail \cite{ERJU} are remarkable examples of this shared aim. In virtually all considered possibilities for such an introduction, the structure of the decision-making process is not questioned: optimization \cite{CacchianiSurvey2014, CormanSurvey2015,  FangSurveyTITS2015, Lamorgese2018}, and recently, machine learning techniques \cite{Khadilkar, GhaNicKirFujHey20, ZhaZha23} are proposed to support dispatchers in their centralized decisions.

A major change of paradigm in this process has recently made the object of a few studies, i.e., a Self-Organizing Traffic Management System (SO-TMS) in which trains autonomously make decisions on their routes and schedules to deal with perturbations \cite{MarPel20:IEEE}. Indeed, self-organizing systems have shown to be effective in many contexts, achieving the most diverse objectives. In the railway traffic management context, self-organization may bring a number of benefits. 
First, thanks to the ability to make local decisions that result in good overall system performance, the system would quickly react to perturbations, thus shortening recovery times. The local decision-making would also allow an efficient scaling up to large networks, while most TMS approaches can hardly do so \cite{X2R-4:D83}.
Second, an SO-TMS may strongly increase the system flexibility, as trains can base their decisions on factors as real-time multimodal traffic conditions or demand flows, possibly disregarding fixed timetables.

Third, it would satisfy the needs and logic of a competitive railway market, as the one envisaged in Europe~\cite{AitAli02122022,BERIA2025100008,freightMarket2025}. In such a market, railway companies compete for the use of the same tracks, which are controlled by infrastructure managers, typically one per country. In this competitive market, companies may not be willing to share information on the value they attribute to specific trains and on the cost of their delay, since it would disclose sensitive data, such as passenger volumes~\cite{CARNEHL2023102902}, to competitors. 
Hence, in a classic centralized system, these costs are unknown by dispatchers and cannot be considered. This lack of knowledge limits decision-making to the plain minimization of  delays.

In a self-organizing system, instead, these known costs do not need to be shared, and trains may negotiate traffic evolutions based on their actual interests. 
Despite these possible benefits, to the best of our knowledge, no such system has ever been convincingly applied to railway traffic management so far. Initial studies on TMS based on train decision-making have been proposed. A first approach proposes that trains decide their movements in a loop-shaped railway line, based on the behavior of the train ahead \cite{shang2018distributed}. Another approach instructs trains to decide on their passing order at a junction based on swarm intelligence principles: trains are grouped in platoons and make decisions as a swarm, by applying rather simple rules after checking for the infrastructure occupation status  \cite{cui2017swarm}. A third approach focuses on resolving conflicts individually by adjusting train schedules or routes \cite{Vanthielen2019}.

Aiming at a more general approach, the SORTEDMOBILITY project~\cite{SORTEDMOBILITY} proposed a formalization of the process through which a SO-TMS may be deployed~\cite{DamNalTibTriPel20}. Specifically, the decision-making process is decentralized and performed by all trains in interaction with each other. First, each train identifies the relevant traffic in its vicinity and generates hypotheses about possible traffic management strategies, relying only on local information. Then, in interaction with neighboring trains, a joint assessment is performed about the feasibility of the proposed strategies, leading to the selection of the best possible strategy to be implemented by each train by means of a consensus process. All the selected strategies are then communicated to the traffic control center, that combines and implements them into a global traffic management plan. This last step is a necessary passage to ensure the consistency and feasibility of the global plan, hence complying with the constraints of real-world operations.

This paper builds upon the framework discussed above \cite{DamNalTibTriPel20} by providing concrete designs and implementations. Specifically, we introduce a spatio-temporal perspective for neighborhood selection, a variant of the state-of-the-art centralized algorithm RECIFE-MILP \cite{PelMarPesRod15:ieeetits} for hypothesis generation, a path-based hypothesis compatibility verification with hypothesis sharing, a voter-model-inspired consensus-seeking approach, and a novel RECIFE-MILP variant for merging selected hypotheses into the global plan.
To the best of our knowledge, this is the first proposal of a complete system based on traffic self-organization that is capable of  quickly reacting to perturbations by consistently producing feasible and optimized traffic management plans in complex networks. 

We thoroughly evaluate our approach on a case study representing traffic in a busy portion of the Italian network, between the cities of Segrate and Ospitaletto in Lombardy, Italy. Here, mixed traffic operated by different railway companies share the available capacity, posing significant challenges related to competition, fairness, and the efficient utilization of limited infrastructure. Our results demonstrate that, in these instances, the proposed self-organizing approach  achieves results that are very close to optimality. Moreover, it  consistently outperforms the state-of-the-art centralized algorithm RECIFE-MILP \cite{PelMarPesRod15:ieeetits},   even when its performance are boosted by the smart reduction of alternative train routes \cite{PasSamPelDarRodPac22:COR}. Specifically, we compare the approaches in terms of total delay suffered by trains in a realistic deployment, exploiting a microscopic railway simulator to reproduce traffic.
In summary, the main contribution of this paper is the proposal of the first comprehensive self-organization approach for railway traffic management. In addition to the formalization and implementation of the algorithms composing this approach, we also propose an articulated experimental analysis to assess its applicability and performance in a possible future deployment. 

The remainder of the paper is structured as follows: Section \ref{sec:litt} discusses related work. Section \ref{sec:rtRTMP} describes the real-time railway traffic management problem. Section \ref{sec:selfOrgProc} describes the self-organization process. Section \ref{sec:selfOrgMod} details the design of each module in our approach. Sections \ref{sec:expSetUp} and \ref{sec:expResults} present the experimental setup and results. Finally, Section \ref{sec:conclusion} concludes the paper and outlines future research directions.

\section{Self-organization and traffic management}\label{sec:litt}
\noindent Self-organization is the process that produces and maintains order and coherence in a system solely through interactions among the system components. Any decentralized system---i.e., a system composed of autonomous agents---can display self-organization if the system components coordinate among themselves without the need of a central authority.
Self-organization is observable in both natural and artificial systems. In Nature, we can think of ant colonies or bee swarms, building complex nests, regulating the division of labor and making collective decisions, always through self-organizing processes \cite{Detrain:2006ev,Seeley:2011bj}. Artificial systems composed of multiple autonomous entities can also self-organize, such as in the case of robot swarms \cite{Dorigo-SR-2021}, cognitive radio networks \cite{LiuEtAl-CR-IEEESMCM-2023} or the Internet of Things \cite{Arellanes-IoT-2021}.

Self-organization is often observed in transportation networks when there is no central authority governing individual movements. There are notable examples in natural systems, which have also inspired solutions for management of vehicular traffic. For instance, ants form effective transportation networks \cite{Garnier:traffic:12}, and use flexible priority rules to regulate traffic on their trails \cite{Dussutour:traffic:13}. Pedestrians, too, self-organize in lanes \cite{Moussaid-pedestrains-2012}, and flexibly adjust their movement and behavior according to the traffic flow \cite{Murakami-pedestrian-adaptive-2021}. The adaptation of control rules according to observed traffic flows is thought to improve vehicle traffic in urban networks, and can be implemented by means of self-organizing traffic lights \cite{Helbing2005,Gershenson2012}. In logistics and in autonomous vehicles, too, self-organization can benefit traffic management with efficient solutions that do not require any centralized traffic regulator \cite{Bucchiarone-2021,Berry-Logistics-2024}. The advantages of self-organization have also been observed in informal public transport networks in Global South, which can be more efficient than centrally organized services \cite{Mittal-SoSouth-2024}. Very often in such systems, agent rules that forego individual benefits for global efficiency prove to be advantageous, resulting in the so-called ``slower is faster effect'' \cite{Gershenson:SIF:2015}.

In railway traffic management, just a few studies address a decentralized formulation of the traffic management problem, as already mentioned in the introduction \cite{shang2018distributed,cui2017swarm,Vanthielen2019}. Various Artificial Intelligence (AI) techniques have also attracted interest in this domain, particularly with the recent advancements in deep neural networks \cite{JusupTrivellaCorman}. For example, agents can improve their decision-making by learning from prior experiences through a reinforcement learning (RL) approach~\cite{Khadilkar}. The Flatland challenges, launched by European railway operators, have encouraged studies in this area \cite{mohanty2020flatlandrl}, providing a simplified railway simulator to evaluate different machine learning (ML) strategies. Nevertheless, implementing learning-based methods in such a complex setting presents difficulties, as the absence of guarantees on solution feasibility, combined with the opaque nature of these models, makes them less appealing to stakeholders.

In summary, the works that propose traffic management approaches in which trains are seen as individual decision makers mostly do so as a  way to decompose the overall problem \cite{Khadilkar, cui2017swarm, Vanthielen2019, JusupTrivellaCorman, mohanty2020flatlandrl}. Here, all trains aim at the objective of minimizing the same function of total delay. When one of those approaches allows, at least in principle, different objectives to be considered by different trains \cite{shang2018distributed, DamNalTibTriPel20}, the corresponding paper either focuses only on a specific part of the traffic management process or reports performance assessments based on very simplified traffic representations. These can consist of a basic network \cite{shang2018distributed}, a minimal set of trains \cite{cui2017swarm}, or a simplified model of train operations \cite{Khadilkar, cui2017swarm, Vanthielen2019, DamNalTibTriPel20, mohanty2020flatlandrl}. Moreover, apart from the paper which we extend~\cite{DamNalTibTriPel20}, only another study considers train decisions that may be critical to traffic feasibility while also ensuring that feasibility is always preserved~\cite{cui2017swarm}.

In this paper, we propose a self-organizing railway TMS developed to address implementation in real-world systems. Our solution focuses on the feasibility of the implementation with respect to current operational requirements. Additionally, the proposed solution is fully explainable as it does not rely on black-box decision models, allowing to precisely identify how and why a certain traffic management decision was taken. This is another relevant aspect to move towards real-world implementation.

\section{Real-time Railway Traffic Management}\label{sec:rtRTMP}
\noindent The real-time railway traffic management problem (rtRTMP) is the problem of choosing train routes and schedules to minimize delay propagation in the event of a disturbance. In particular, all trains traveling in a predefined portion of the railway network (\textit{control area}) are considered. Their potential \textit{conflicts} are identified and resolved, where a conflict is an event in which two or more trains, if traveling at the planned speed, would be utilizing the same track concurrently: such an event implies unplanned brakings or stops to preserve safety distances. When a microscopic infrastructure representation is considered, as in this paper, the track portions in which conflicts are detected are the \textit{track detection sections} (TDS), which are the smallest portions that allow train presence detection in reality. By considering such microscopic representation, including details on train movements and interlocking systems \cite{pachl_railway_2002}, the solutions produced when solving the rtRTMP can be directly deployable.

In this paper, we consider a deployment in which a \textit{closed loop} is enacted between the field and the decision making process \cite{Vanthielen2019,QuaPelGovAlJaeMarRodDolAmbCarGiaNic16:TRC,CorQua15}, be it self-organized or centralized. Here, the traffic evolution is controlled through the Real-Time Traffic Plan (RTTP) \cite{QuaPelGovAlJaeMarRodDolAmbCarGiaNic16:TRC}, which indicates train routes  and passing orders
.  This evolution is constantly monitored and, periodically, reported in the form of a Traffic State (TS) \cite{QuaPelGovAlJaeMarRodDolAmbCarGiaNic16:TRC}.  When a TS is available, the decision making process is executed and produces a new RTTP to react to the possibly observed perturbations.

\section{Self-organization process}\label{sec:selfOrgProc}
\begin{figure}[!tb]
\includegraphics[width=\columnwidth]{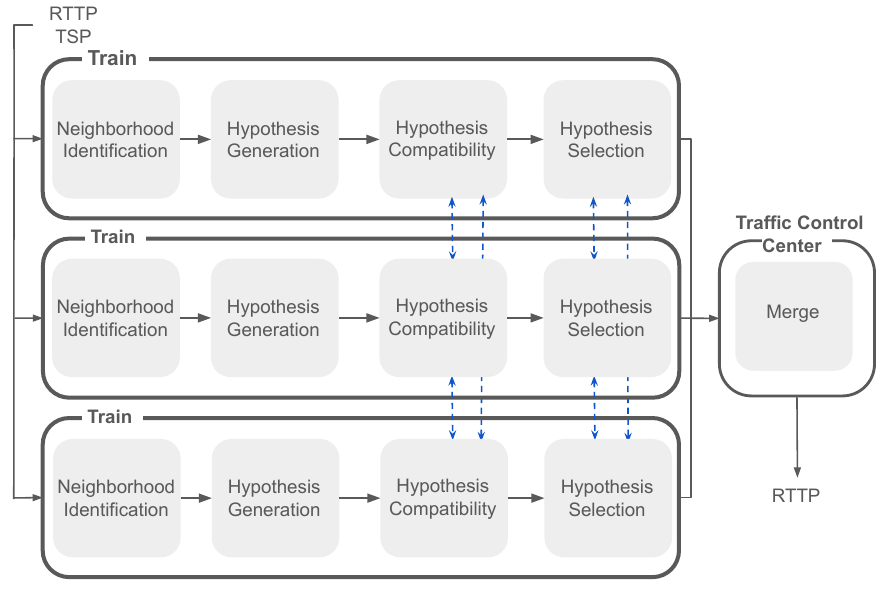}
\caption{\label{fig:pipeline}Conceptual representation of the self-organized traffic management process \cite{DamNalTibTriPel20}.}
\end{figure}

\noindent To self-organize, trains need to identify the traffic share with which they may interact in the near future, then to generate hypotheses on how these shares may evolve, to finally assess other trains' hypotheses and get to an agreement on which ones shall be implemented. This process has been formally defined in \cite{DamNalTibTriPel20} through five modules (see Figure~\ref{fig:pipeline}).
When trains receive the current RTTP and TS, they independently carry out a \textit{neighborhood identification} to determine the trains they could be in conflict with in the near future, and that therefore need to be taken into account for traffic management. Then, each train executes the \textit{hypothesis generation} to formulate possible traffic management strategies involving all its neighbors. Subsequently, it shares these hypotheses with its neighbors and carries out a \textit{hypothesis compatibility} check to verify which of the hypotheses received from its neighbors could be selected concurrently with its own without generating feasibility issues. 
In the last step carried out independently by each train---the \textit{hypothesis selection}---each train engages with its neighbors in an iterative process, during which they all try to reach an agreement on the hypotheses to select, minimizing their own costs while preserving overall compatibility with neighbors.
Finally, the trains share their selected hypotheses with the traffic control center, which is in charge of the \textit{merge} of all hypotheses into an RTTP. If incompatibilities are detected at this stage, which may specifically happen for trains that are currently too far to be included in each other neighborhoods, the control center repairs the RTTP by resolving the detected conflicts without modifying the selected hypotheses in the short term.

\begin{figure}[!tb]
\centering
\includegraphics[width=0.65\columnwidth]{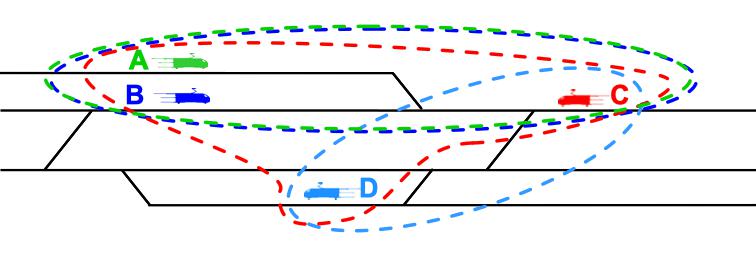}
\includegraphics[width=\columnwidth]{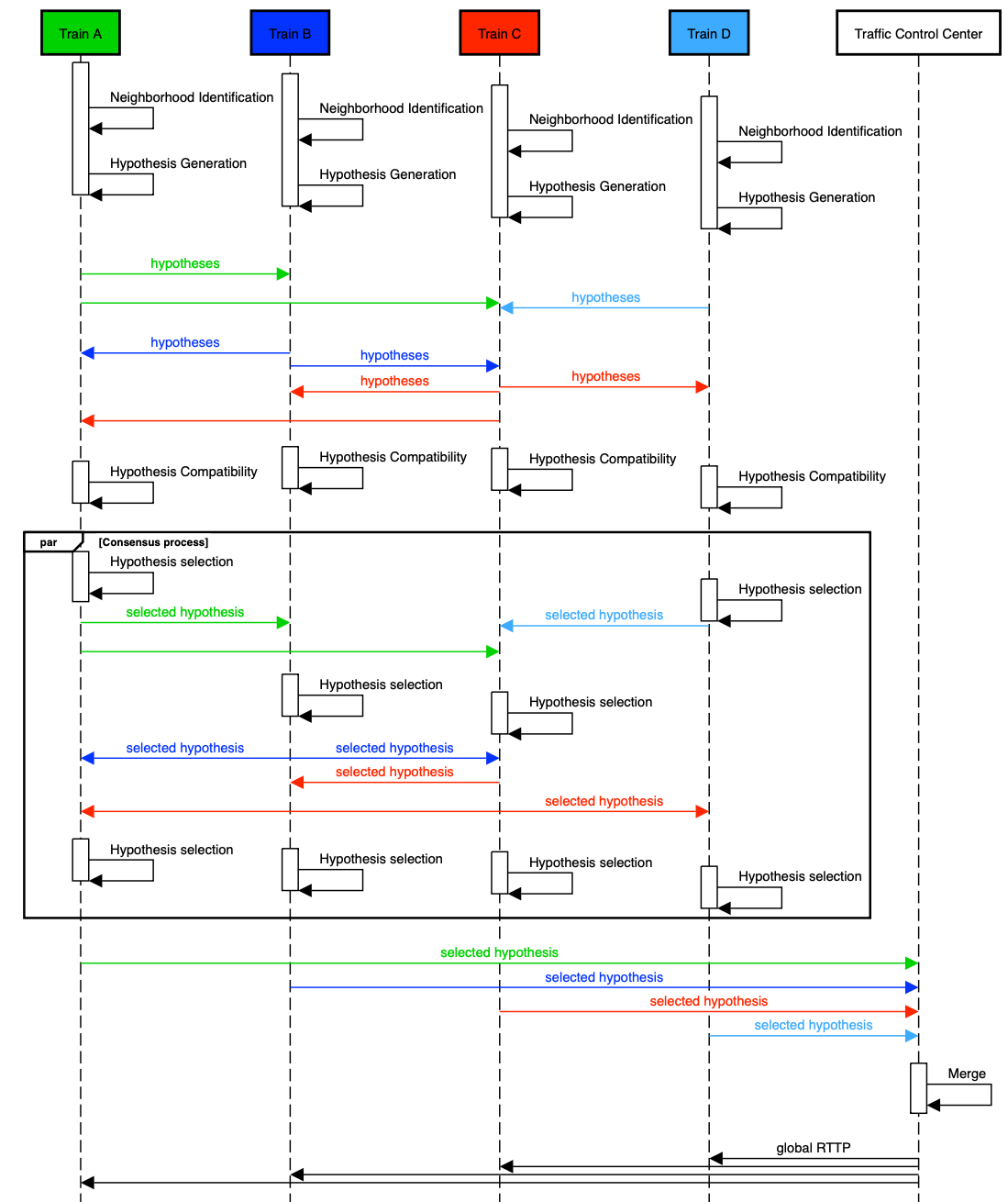}
\caption{Top: Four trains traveling in a control area. Each train's neighborhood is indicated using the corresponding color. Bottom: sequence diagram of the self-organization process with interactions and control processes for each train \cite{DamNalTibTriPel20}.}
\label{fig:example}
\end{figure}
Figure~\ref{fig:example} depicts an example of its application. Here, four trains travel in a control area. After neighborhood identification, A and B have the same neighborhood (A, B and C), C includes all trains in its neighborhood (A, B, C, D) while D only has C in its neighborhood (D, C). Then, each train generates traffic management hypotheses and sends them to its neighbors for the compatibility check. Subsequently, a hypothesis selection process takes place whereby each train iteratively chooses the preferred hypothesis depending on its corresponding cost and its compatibility with the hypotheses selected by its neighbors. After this selection, the chosen hypotheses are sent to the traffic control center and merged in the new RTTP.

\section{Self-organization module designs}\label{sec:selfOrgMod}
\noindent In this section, we detail the design we propose for each module described in Section~\ref{sec:selfOrgProc}. In the following, we refer to the train that actively engages in the decision making process as the \textit{focal} train.

\subsection{Neighborhood selection}\label{subsec:neighSel}
\noindent The first step of the self-organized railway traffic management is the identification of neighboring trains within the network. A neighbor of a focal train $t$ is defined as any train $t'$ that could potentially conflict with $t$ within a specified time horizon $T_h$, starting from the current time. Our proposed algorithm first performs a prediction of traffic evolutions based on the current TS and RTTP. Then, to identify its neighbors, each train exploits a graph representation of the rail network, in which nodes correspond to the TDSs and links describe connections among them. Each TDS is associated with the IDs of the trains that may occupy it in the future---according to all possible routes that a train can take to reach its destination---along with their occupation times and route information. Possible routes and occupation times can be predicted for any train, as the former depend on the train and infrastructure characteristics, and the latter from the quite standard application of driving practices. Predicted occupation times are calculated by adding the running times for each TDS on the train's route to the current time. If a train has multiple feasible routes passing through the same TDS, then all possible occupation times are considered. After completing the mapping of trains to TDSs, each train identifies the TDSs it can occupy within the time horizon $T_h$ (i.e., the expected occupation time falls between the current time $T$ and $T+T_h$). Any other train that may occupy the same TDS within $T_h$ is marked as a neighbor.
Rolling stock reutilization constraints are also considered: once a train is included in a neighborhood, all other trains linked to it by a rolling stock reutilization constraint are also included. 
Specifically, any group of trains sharing rolling stock is treated as it was a single focal train, resulting in a single neighborhood that merges the neighbors of all trains in the group.

As a result of the neighborhood identification process, each train $t$ has a set of other trains $\mathcal{N}(t)$ with which it should coordinate. At the global level, we can represent these neighborhoods by means of an \textit{interaction graph} $G_{I} = (\nu_{I} , \epsilon_{I})$. Here, nodes in $\nu_{I}$ represent trains, and edges in $\epsilon_{I}$ represent neighborhood relations: an edge connects two nodes if the corresponding trains are in each other's neighborhood.

\subsection{Hypothesis generation}\label{subsec:hypGen}
\noindent To generate different hypotheses of traffic management strategies, each train $t$ considers its own neighborhood and tries to optimize schedules and routes to minimize delays. To this end, a variant of the RECIFE-MILP algorithm for the rtRTMP \cite{PelMarPesRod15:ieeetits} is employed, first proposed in \cite{DamNalTibTriPel20}. 
In particular, this algorithm is based on the solution of a mixed-integer linear programming (MILP) formulation and it operates in two phases. First, train schedules are optimized only considering the routes of the input RTTP. Second, this solution is used as a warm start and all route alternatives are considered. In the employed variant, routes and schedules are optimized only for the focal train $t$ and its neighbors $t'\in\mathcal{N}(t)$. The paths of all the other trains are considered as hard constraints: the routes must remain the ones of the  input RTTP,  and the schedules may be changed provided that the trains exit the control area no later than what expected before the optimization. The algorithm stops as soon as a solution is proven optimal or when the available computational time has elapsed.

%

The objective function minimized is the weighted sum of train delays at their exit from the control area. The weights represent the cost of delays  of the focal train for itself and all the other trains. By considering different values for these weights, which are an input of the algorithm, this objective function can model individual train behaviors in different contexts. In particular, it can model extreme competition or extreme cooperation, or any combination of them~\cite{RICCIARDI2022320}.  For example, each train may be interested in minimizing only its own delay, and it  may set its own weight to a positive value, while the weights for all the other trains are zero. Alternatively, each train may value its own delay the most, but it may also wish for a well-performing railway system as a whole. Hence, it may set its own weight ten, for example, times higher than the one for the other trains, but keep all weights strictly greater than zero. Moreover, this objective function can capture the lack of information a train may have on the cost of delay for its competitors~\cite{CARNEHL2023102902}. In this case, the weight may represent an educated guess on the other trains' cost rather than a true value. This is the rationale we adopt to define the costs in the computational analysis (Section~\ref{sec:expSetUp}).
 
However, let us remark that any linear function of delay may be used in different scenarios.

The considered RECIFE-MILP variant produces a number of different solutions. 
Each solution computed by train $t$ is a hypothesis $h_t$.  It includes a path for each train, where a path describes the train utilization of all the TDSs along its route: it can be either the original or a new path for the focal train and the trains in its neighborhood, while it is necessarily the path of the input RTTP, or an extremely similar one, for all the other trains. All these paths describe a complete RTTP, and, as such, all train routes and passing orders.  At most $\bar h$ solutions are retained, those that are not further than $p\%$ from the optimal one or the best one identified by the algorithm, with $p$ and $\bar h$ input parameters.
We refer to the set of hypotheses generated for train $t$ as $H_t$.

After running the RECIFE-MILP variant, the hypothesis generation can include the \textit{hypothesis sharing}: each train $t$ receives from each neighbor $t'$ its hypotheses, and it retains all those that do not correspond to its own. Then, the corresponding cost is computed according to the private information available to train $t$. For avoiding an explosion in the number of exchanges, only hypotheses generated by applying the RECIFE-MILP variant are shared: trains do not share hypotheses they receive from other neighbors. 
By applying this sharing, each train can increase the number of its hypotheses at virtually no computational effort, with the additional benefit of these hypotheses being certainly compatible with the ones originally shared by neighbor trains. 

\subsection{Hypothesis compatibility}\label{subsec:hypComp}
\noindent A pair of hypotheses $h_1$ and $h_2$ generated by two neighboring trains $t_1$ and $t_2$ may differ in multiple aspects related to scheduling and routing.
However, despite differences, they may contain  paths  for $t_1$ and $t_2$ that, if implemented, produce no infeasibility issues. In this sense, $h_1$ and $h_2$ are said to be compatible. 
Specifically, to perform the compatibility check, the  paths  of $t_{1}$ from $h_{1}$ and the one of $t_{2}$ from $h_{2}$ are confronted. If there exists at least one TDS in which utilization times  overlap among $t_1$ and $t_2$, then $h_{1}$ and $h_{2}$ are declared incompatible.

After the compatibility checks, each train $t$ is aware of the compatibility of any hypothesis $h\in H_t$ with those of the neighbors. At the global level, we can represent all compatibilities in the so-called \textit{hypothesis graph} $G_{H} = (\nu_{H} , \epsilon_{H})$, where $\nu_{H}$ represents all hypotheses from all trains and $\epsilon_{H}$ denotes the edges connecting compatible hypotheses. To each node of $\nu_{H}$, a cost is associated as determined by the hypothesis generation process. Note that $G_{H}$ is a $N$-partite graph, where $N$ is the number of trains in the system: edges exist only among hypotheses of different trains.

\subsection{Hypothesis selection}\label{subsec:consensus}
The selection of hypotheses by each train is achieved through a self-organized consensus process, that is, an iterative procedure inspired by voter-like models in which each train $t$ selects one candidate hypothesis $h_t\in H_t$ in the attempt to maximize compatibility with the selection made by neighboring trains. Coordination is achieved through interactions among trains trying to find a feasible traffic management plan, using only local information about their neighbors.
At the global level, this process aims to reach a (possibly optimal) solution to a given problem instance $(G_I, G_H)$, where the goal is to find a sub-graph $G_H'$ that assigns a single hypothesis to each train, minimizing the overall cost. 

The self-organization process starts with each train selecting its lowest-cost hypothesis as its initial state. At each subsequent iteration, train $t$ can decide either to keep the current selection $h_t$ or to select a different hypothesis $h_t' \in H_t$. This choice depends solely on the local information available to the train, that is, the degree of compatibility of $h_t$ with the hypotheses currently selected by its neighbors. It is important to note that the cost of neighbors' hypotheses is not known, as this information is private to each train.

To evaluate the level of compatibility,  train $t$ randomly selects up to $K$ trains from its neighborhood $\mathcal{N}(t)$, observes their current hypotheses (e.g., via directly querying the neighbors), and ranks all hypotheses $\hat{h} \in H_t$ according to the compatibility count $\chi(\hat{h})$, i.e., the number of the $K$ selected neighbors' hypotheses that are compatible with $\hat{h}$.

As discussed below, hyperparameter $K$ controls the trade-off between trying to find a feasible solution quickly and aiming at a high-quality solution from each train's perspective.

Train $t$ then decides whether to keep its current hypothesis or switch to a more compatible one according to the following policy:
\begin{enumerate}
    \item If $\chi(h_t) = K$, i.e., the selected hypothesis $h_t$ is compatible with all the chosen neighbors' hypotheses, then the train keeps $h_t$ as its current hypothesis.
    \item If $\chi(h_t) < K$, then train $t$ selects a more compatible hypothesis $h'$ from the ranking, so that $\chi(h') = \max_{\hat{h}} \chi(\hat{h})$. If there are multiple hypotheses with the same compatibility count, the train chooses the one with the lowest cost.
\end{enumerate}
This default policy allows the trains to gradually adjust their hypotheses towards a configuration where all neighboring trains hold compatible hypotheses, while prioritizing hypotheses with the lowest cost.
The iterative process continues until one of the following conditions is met: (i) all trains achieve a state where their hypotheses are compatible with all their neighbors, i.e., consensus is reached; or (ii) a predefined maximum number of iterations is performed, resulting in a failure in achieving global consensus. Note that the interaction graph may be disconnected, presenting multiple connected components, i.e., sub-graphs in which starting from each node it is possible to reach any other node. In this case, it is possible that consensus is reached within a given connected component but not in another one.

As anticipated above,  the hyperparameter $K$ tunes a speed-accuracy trade off. High values of $K$ imply that the train considers multiple (possibly all) neighbors during the decision making process. This can lead a train to simultaneously seek compatibility with a large set of neighbors, hence increasing the speed of convergence towards a shared solution. However, in such conditions, it may be difficult to find a compatible hypothesis, and costs may increase. Conversely, when $K$ is small (and possibly, $K = 1$), the train considers only a small set of neighbors at a time, and therefore the speed of convergence may be slower. However, in these conditions, a train has a higher chance of selecting among multiple compatible hypotheses, and convergence to low-cost hypotheses can gracefully build up within the network.
To maximize decision accuracy and speed, we propose an adaptive strategy that adjusts the value of the hyperparameter $K$ over time. Trains start the consensus process with the highest possible value of $K$ (i.e., $K$ equal to the total number of neighbors) and gradually decrease it towards $K=1$ from iteration to iteration. In this way, the self-organized process allows to rapidly move towards agreement with neighbors, and to gradually improve decision quality once trains start being able to prioritize hypotheses with smaller cost. In \cite{d2025decentralised}, we extensively investigated the role of $K$ and compared various strategies, concluding that the adaptive approach features both the highest convergence speed and the highest rate of convergence to an optimal solution.

Algorithm~\ref{alg:cons} summarizes the hypothesis selection procedure from the perspective of a focal train.

\begin{algorithm} [!tbp] 
\caption{Hypothesis selection procedure of each focal train $t$}
\begin{footnotesize}
\label{alg:cons}
\begin{algorithmic}[1]
\Statex \textbf{Input:} Hypothesis graph $G_{h}$.
\Statex \textbf{Returns:} Each train's selected hypothesis.

\State $h_t \gets t.\mathit{selectMyMinimumCostHypothesis}()$
    \label{line:hypothesis-selection-start}
\While{ $\exists$ neighor with selected hypotheses incompatible with $h_t$\\ \hspace{0.8cm}\& available time not elapsed }
   \State  $t.\mathit{updateNeighborSelectedHypotheses}()$
   \State ${\Tilde{\mathcal{N}}(t)} \gets t.\mathit{selectNeighborSubset}(\mathcal{N}(t))$
   \State $t.\mathit{rankHypotheses}(H_t, {\Tilde{\mathcal{N}}(t)})$
    \If{$\chi(h_t) < K$}
      \State $h_t \gets t.\mathit{selectNewHypothesis}(\boldsymbol{\chi})$
    \EndIf
    \State $t.\mathit{shareHypothesisWithNeighbors}()$    
\EndWhile \label{line:hypothesis-selection-end-while}
\State \Return $h_t$
\end{algorithmic}
\end{footnotesize}
\end{algorithm}

\subsection{Merge}\label{subsec:merge}
\noindent After the conclusion of the consensus process, each train proposes its selected hypothesis to the traffic control center. Here, the merging phase begins to obtain a new RTTP starting from the input RTTP. First, the traffic control center considers the connected components of the interaction graph in which consensus was reached. For each train $t$ belonging to one of these connected components, the path in the input RTTP is replaced with the one of the selected hypothesis $h_t$. For the trains in connected components that did not reach consensus, the path of the input RTTP remains unchanged. 

Once this merging is performed, a consistency check is carried out on the updated RTTP to identify TDSs where paths overlap, if any. 

Indeed, when hypotheses are generated, each focal train $t$ considers the paths of trains not belonging to its neighborhood (e.g., $\bar t$) as constraints. If these paths are not changed by the respective trains by means of their hypothesis selection, the new path of the focal train~$t$ will not raise any inconsistency. However, if $\bar t$, as a focal train itself, achieves a consensus with its own neighbors and changes its path, there is no guarantee that the new choice of $t$ and $\bar t$ will be consistent. If two or more paths overlap, 

a repair procedure is executed exploiting a variant of the RECIFE-MILP algorithm \cite{PelMarPesRod15:ieeetits}, in which train routes are preserved, as well as train schedules for which a consensus was found. In particular, this is the case for all scheduling decisions that are realized within the time horizon $T_h$ of the neighborhood selection. All other scheduling decisions are optimized, considering the minimization of the unweighted total delay. In fact, the control center has no access to the weights assigned by the railway companies to their trains.

Remark that the fulfillment of the merging phase at the control center departs from typical, strictly self-organizing processes, where all decisions are based solely on local information. However, a global check is the only option to guarantee the overall feasibility of the RTTP. As we are proposing here an approach which may in principle be deployed in practice, supplying an overall feasibility guarantee is a necessary condition. This global check could be performed by a single train, but we posit that the traffic control center is a more sensible choice to ensure the fairness of the repair operation, if necessary, and remain to some extent consistent with the current traffic management organization.

\section{Experimental setup}\label{sec:expSetUp}
\noindent To assess the performance of the proposed SO-TMS and compare it against baseline approaches, we set up a closed loop experimental framework that connects the OpenTrack microscopic railway simulator \cite{NasHur04} with the TMS through an API that regulates the information exchanges between the two programs, as well as synchronization. The OpenTrack simulator and the TMS exchange information exploiting the RTTP and TS formats as explained in \cite{DamNalTibTriPel20} and illustrated in \cite{RTTP-TS}. Specifically, the TS lists the trains that either are currently in the control area ({\it{in-area}} trains) or are predicted to enter it ({\it{out-of-area}} trains) within a fixed horizon from the current time~(40 minutes in this paper). 
For each in-area train, the TS indicates which TDS the train's head is currently occupying and the timestamp at which it was first detected there. For each out-of-area train, the TS indicates the expected entry timestamp. The~RTTP reports, for each train, the sequence of TDSs that the train is scheduled to traverse and the time at which it is predicted to do so. From these times, passing orders can be obtained. In our implementation, the data encoded in TS and RTTP are exchanged via xml files.

\begin{figure*}
    \centering
    \includegraphics[width=\textwidth]{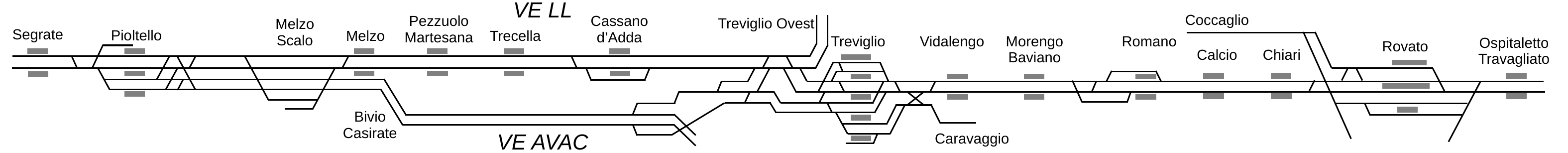}
    \caption{Schematic representation of the Segrate-Ospitaletto control area}
    \label{fig:PiolRov}
\end{figure*}

We consider a control area corresponding to the $60$ km line connecting Segrate and Ospitaletto Travagliato, in Italy, as represented in Figure~\ref{fig:PiolRov}. Two lines branch off from the Pioltello station:
the Pioltello--Treviglio--Rovato (VE LL) line which from Pioltello reaches Rovato, encountering a large number of stations and stops; 
the Pioltello--Bivio Casirate--Treviglio (VE AVAC) line which runs parallel to the VE LL line up to Melzo Scalo, and then enters the Milan-Brescia highspeed line at the Casirate junction. From Casirate, through an interconnection, the line also branches off to Treviglio. About $150$ trains per day circulate in this control area. Each train can use between $2$ and $64 \, 512$ alternative routes, with a mean and median of $9 \, 366$ and $256$, respectively.

Here, freight, intercity, regional and high speed trains belonging to different companies share the same tracks. No detailed information is available on the historic train loads, be them passenger or freight trains, nor on the company operating each train. To mimic the competitive context, we assume that each train is independent from the others. Each train knows its own cost of delay, and makes an educated guess on the one of the other trains. This cost is the weight used in the objective function for the hypothesis generation process (Section~\ref{subsec:hypGen}). 
 The choice of having all independent trains does not precisely capture reality, where many trains are operated by the same companies. However, we conjecture that this does not have a major impact on the results, as we assume trains aim at operating in an overall well-performing system by considering all trains' delays in their objective functions.  %
In agreement with the industrial partners of the SORTEDMOBILITY project \cite{SORTED:D52}, weights depend on the types of trains: for freight trains, weights vary in the interval
$[15, 24]$; for regional trains, in the interval $[21, 36]$; for intercity trains, in
the interval $[19, 27]$; for high-speed trains, in the interval $[24, 40]$; for empty rides of passenger trains, in the interval $[15, 18]$.  From these intervals, we randomly select the weight for each train. We consider a uniform probability distribution inspired by what has been done in the literature for assigning capacity to competing railway companies during timetable production \cite{BornGroLukMitSchSchTan:2006}. 
%
%
%

\begin{figure}
    \centering
    \includegraphics[width=\columnwidth]{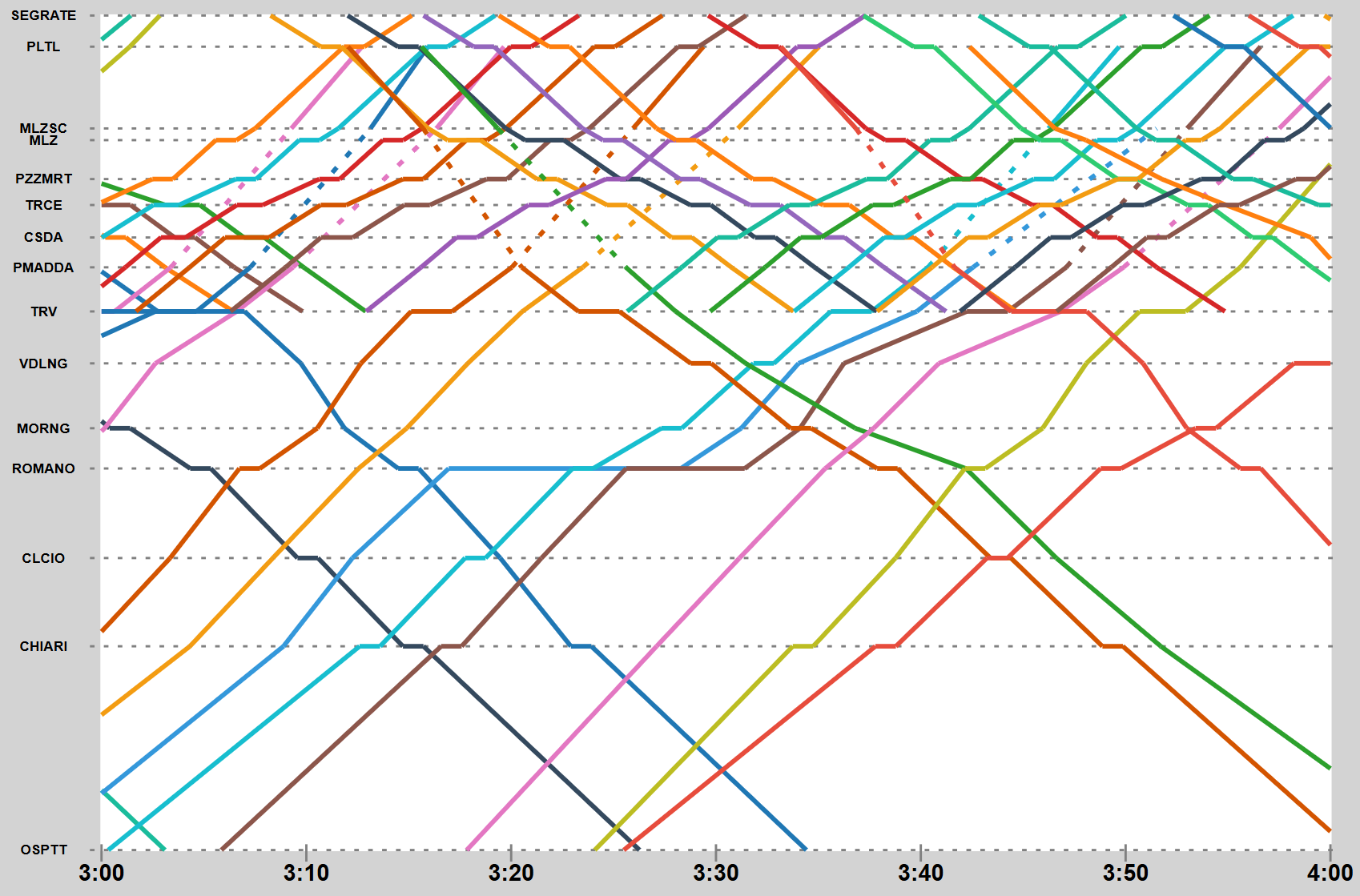}
    \caption{Space-time diagram representing the timetable considered.}
    \label{fig:timetable}
\end{figure}

For obtaining the traffic scenarios to be tackled, we apply two steps: first, we compress the timetable actually operated on this control area; second, we create random perturbations based on historical observations. We compress the timetable because, in a preliminary analysis reported in \cite{SORTED:D52}, we realized that no significant margin for optimization existed in the original one. To compress the timetable, we consider the scheduled train path of each train, and we shift it to the earliest possible time such that: (i) no path overlap exists; (ii) the passing orders from the original timetable are preserved; (iii) no braking is necessary due to traffic;  (iv) each planned stop is as short as possible, but it is never shorter than 30 seconds, which is the minimum dwell time; and (v) the total path duration is as short as possible. By doing so, we obtain a very dense mixed traffic situation, one hour of which is represented in a space-time diagram in Figure~\ref{fig:timetable}. The full-day real timetable takes 7 hours to be operated after the compression, in nominal conditions.
Here, each colored line is a train path, the horizontal lines correspond to stations along the line. Dashed lines show the path of trains using the VA AVAC branch in Figure~\ref{fig:PiolRov}, between Bivio Casirate and Treviglio.
%
%
To make instances more tractable, we partition each scheduled train in the timetable into a sequence of shorter trains: the shorter trains use the same rolling stock, and they are linked by re-utilization constraints at stations. In particular, a single train going from origin to destination through multiple stops is represented by multiple trains doing just one stop: one going from the origin to the first stop, one from the first to the second stop, and so on, up to the last one going from the last stop to the destination. Between pairs of subsequent trains in these sequences, compatibility constraints state that the TDS where the first train arrives must be the same from where the second train departs, and that the second train cannot depart before the first one has arrived plus a minimum rolling stock reutilization time, which here corresponds to the minimum dwell time. The so-obtained timetable is characterized by a significantly higher number of trains, but with a reduced number of alternative routes: it includes $456$ trains having between $1$ and $44$ alternative routes, with mean and median of $6$ and $4$, respectively.
Partitioning the original trains makes the instance more tractable for RECIFE-MILP, which defines one binary variable for each train-route pair. Indeed, when trains pass stations with multiple alternative platforms, their number of routes grows combinatorially. Instead, partitioned trains that pass through fewer stations have fewer routes, and the number of variables grows at a much slower pace. For example, a train passing four stations, each with three platforms, will require 81 binary variables, while only 27 after being partitioned (three trains with nine routes each). In the neighborhood selection, partitioned trains coming from the same original one are considered jointly: as soon as one of them is in a neighborhood, all the others are also included. By doing so, the only impact of the partitioning is the speed-up of RECIFE-MILP.

\begin{table}
    \centering
    \caption{Parameters of the probability distributions used to draw random entrance delays in the perturbation scenarios, depending on train type and entrance location. Intervals are indicated in minutes.}
    \label{tab:perturbations}
    \begin{tabular}{l|r|r|r|r|r|r}
Interval: &0 & (0,5]& (5,15] & (15,30] & (30,60] & (60,180]\\
\hline
\multicolumn{7}{l}{From Ospitaletto Travagliato}\\
Passenger & $0.07$ & $0.65$ & $0.21$ & $0.04$ & $0.01$ & $0.02$ \\
Freight & $0.31$ & $0.05$ & $0.11$ & $0.15$ & $0.14$ & $0.24$ \\
\\
\multicolumn{7}{l}{From Segrate}\\
passenger & $0.24$ & $0.47$ & $0.24$ & $0.03$ & $0.01$ & $0.01$ \\
freight & $0.28$ & $0.07$ & $0.10$ & $0.11$ & $0.17$ & $0.27$ \\
    \end{tabular}
\end{table}

In our experiments, we consider $35$ perturbed traffic scenarios that include 93 trains. These are all trains entering between \mbox{1:00} and \mbox{4:00}~AM in the compressed timetable. Looking at the original timetable, this time span covers all trains entering the control area before \mbox{3:15}~PM. In each perturbed scenario, trains enter the infrastructure with a random delay, and any additional change to the plan is a consequence of the entrance delays and the resulting traffic congestion (e.g., the need to reduce train speeds or to change dwell times at stations). Such entrance delay perturbations are very frequently considered in the literature on the rtRTMP \cite{PelMarPesRod15:ieeetits,Caimi2012,CORMAN201271}. The distributions from which entrance delays are drawn are deduced from historical data. They are piecewise uniform distributions, summarized in Table~\ref{tab:perturbations}. They depend on the type of train and on the entrance point. 
In the table, we indicate the probability that each train has a delay in the interval heading each column.  For each train, we draw a random number between zero and one, and we identify the piecewise distribution corresponding to its type and entrance point. If the number drawn is smaller than or equal to the end of the first step (e.g., 0.24 for a passenger train from Segrate), then we assign no entrance delay to the train. If it is between the beginning and the end of the second step, we assign a delay equal to the corresponding value in this step. For example, if the random number drawn for a passenger train from Segrate is 0.475, we first identify the corresponding step, i.e., the second one that corresponds to a delay between zero and five minutes. Then we observe that it falls right in the middle of this step ($0.475-0.24=0.237$, which is $0.47/2$), and we assign a delay equal to the middle of the corresponding interval to the train (2.5 minutes).

Depending on the perturbation, the total duration of scenarios exploiting SO-TMS varies between $4$ hours and $30$ minutes and $6$ hours: the exit time of the last train from the control area determines the end of the scenario, and this depends on the specific trains that suffer delays. 
In the closed loop, decision making by any TMS is triggered every five minutes. Depending on the duration of each scenario, with SO-TMS decisions are made between $56$ and $74$ times. 
When triggering decision making, a TS is produced, including all focal trains and all trains present or entering the control area within the next $40$ minutes.

With this framework, we compare the performance of  five  traffic management approaches:
\begin{itemize}
    \item SO: the decision making is performed though the self-organization process and modules described in Sections~\ref{sec:selfOrgProc} and~\ref{sec:selfOrgMod};
    \item CEN-1: the decision making is carried out centrally through the RECIFE-MILP algorithm \cite{PelMarPesRod15:ieeetits} in a real-time context;
    \item CEN-2: the decision making is carried out by using the ACO-TRSP algorithm \cite{PasSamPelDarRodPac22:COR} to limit the number of alternative routes available for each train, and by chosing routing and passing orders through the RECIFE-MILP algorithm \cite{PelMarPesRod15:ieeetits} in a real-time context;  
    \item CEN*: the decision making is carried out centrally through the RECIFE-MILP algorithm \cite{PelMarPesRod15:ieeetits} allowing a very long computational time, aiming to find optimal solutions; 
    \item FCFS: traffic evolves following the \textit{first come first served} principle.
\end{itemize}
In SO, given the current TS, the neighborhood selection uses a time horizon of $T_h = 15$ minutes, and each train generates at most two hypotheses: the optimal one, or the best one the hypothesis generation found within the time limit, and the one corresponding to the current RTTP. The time limit for each run of the hypothesis generation is set to three minutes. Hypothesis generation considers the weighted sum of the delays as the objective function, and the generated hypotheses are shared among neighboring trains. Finally,  in our simulation of the decentralized hypothesis selection process, we simulate asynchronous train decisions by running sequential iterations, in each of which we randomly select one train among the focal ones to make a decision. The time limit is represented by a maximum number of iterations, which we set to $100\ 000$. If no feasible solution is found after these iterations, 
a consensus failure is reported. 

In CEN-1, the optimization is carried out minimizing the sum of delays, hence the weights are not considered as we assume this is private information available only to trains. For each execution  of the optimization, the same time limit of three minutes is considered to solve instances including all trains in the TS. This setup mimics the one of previous studies on this type of deployment \cite{CorQua15, QuaPelGovAlJaeMarRodDolAmbCarGiaNic16:TRC}.

In CEN-2, we use the same setting as in CEN-1, but, at each execution, 30 seconds are used by ACO-TRSP to find the six most suitable alternative routes per train, and two minutes and 30 seconds are available for RECIEF-MILP, as in the reference paper \cite{PasSamPelDarRodPac22:COR}.

In CEN*, we use the same setting as in CEN-1, but, at each execution, we let RECIFE-MILP run for one hour.

In FCFS, a myopic strategy is enacted, so that whenever a conflict arises, the train that can pass first has priority, and the others follow.

The OpenTrack simulations are run on a Windows~10 13th Gen Intel Core~i7 13700 CPU @ 2.10 GHz and 64~GB of RAM, while the decision making on Ubuntu 18.04.6 LTS with an Intel Xeon CPU E5-2637 v3 @ 3.50GHz and 128~GB of RAM. RECIFE-MILP and its variants 
are coded in C++ and use IBM ILOG CPLEX v12.6 as the MILP solver, while all other modules are coded in Python.

\section{Experimental results}\label{sec:expResults}
\noindent 
In this section, we analyze the performance of the proposed TMS, comparing the results obtained with SO, CEN-1,  \linebreak CEN-2, CEN*  and FCFS. 
Each SO~module requires the following minimum, average, and maximum computational times per execution (in seconds): neighborhood selection - 5, 12 and 23; hypothesis generation - 5, 28 and 180; hypothesis compatibility - 0, 1 and 5; hypothesis selection - 1, 2 and 2; merge - 5, 6 and 32.

The metric we use to assess the performance of SO is the percentage improvement with respect to  each  baseline approach, i.e., CEN-1, CEN-2, CEN*  or FCFS. In the following, we will refer to it as \textit{improvement}, for brevity.
We compute this improvement for both total weighted and unweighted delay over the $93$ trains of each scenario. Specifically, we consider the delay with which they exit the control area in simulation, with respect to the timetable. In the computation of the weighted delay, we use the weights that are considered by trains in their own hypothesis generation in SO. Hence, this total weighted delay represents the actual total cost of delay.

Note that, while the total unweighted delay corresponds to CEN-1's, CEN-2's and CEN*'s objective function, the total weighted one is never used to optimize: in SO no train knows the exact weight of the others and can only use educated guesses. Nonetheless, we think these two indicators are appropriate for our analysis, as they are the ones that best represent the overall performance of traffic management.

\begin{figure}[!b]
    \centering
    \includegraphics[width=1\linewidth]{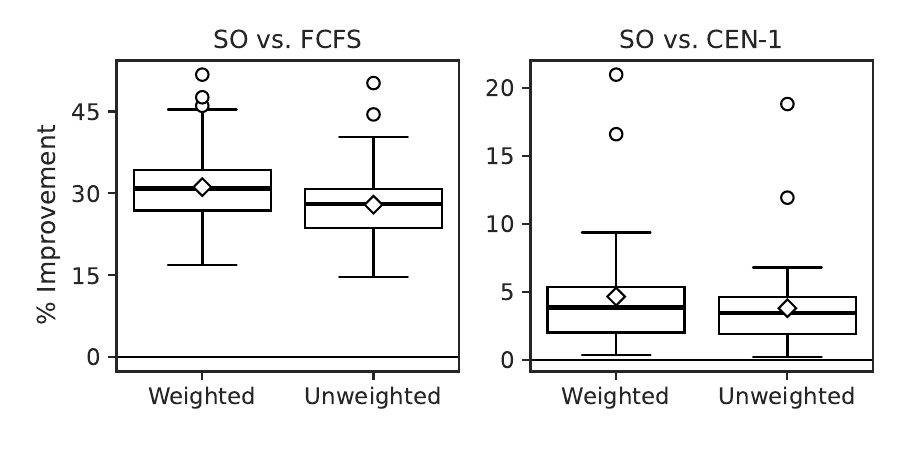}
    \includegraphics[width=1\linewidth]{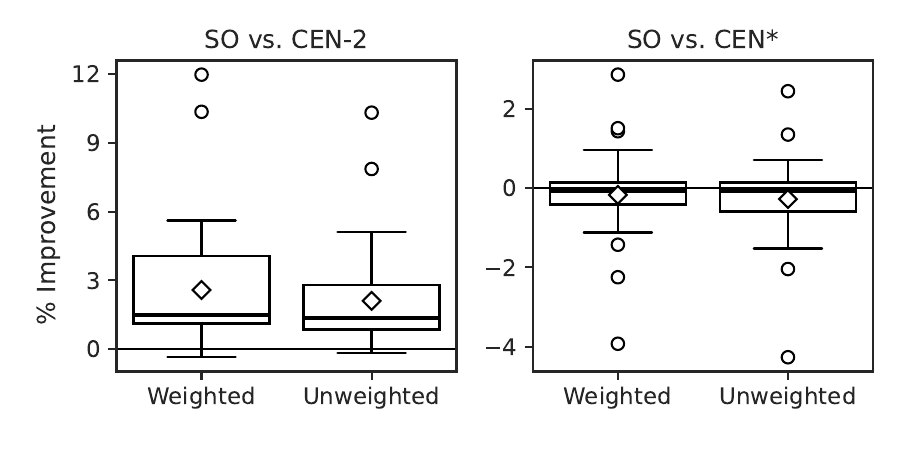}
    \caption{Percentage improvement of total weighted and unweighted delay of SO w.r.t FCFS, CEN-1, CEN-2 and CEN*.}
    \label{fig::improvement_vs_centralized}
\end{figure}

Figure~\ref{fig::improvement_vs_centralized} reports the distribution of the improvement of SO over  the four benchmarks  across all scenarios. It is clear that SO outperforms FCFS, CEN-1 and CEN-2: for the first two, all points of the distributions are strictly above zero, meaning that SO never produces higher delays than the other approaches. For CEN-2,  SO is worse than this benchmark in two out of the 35 scenarios, by $0.1$ and $0.2\%$, and  $0.2$ and $0.3\%$ for unweighted and weighted delay, respectively. In the remaining 33 scenarios, SO regularly improves the solution by a few percentage points.  
The average improvements are represented by diamonds in the figure.  In the SO vs. FCFS comparison, the average improvement is about $31\%$ for the weighted delay, and about $28\%$ for the unweighted one. Given the already good performance of the centralized benchmarks, for SO vs \linebreak CEN-1, these values decrease to $5\%$ and $4\%$, respectively. Since, in these experiments, CEN-2 performs better than \linebreak CEN-1 due to the large number of route alternatives in the original instances, the improvement brought by SO when compared to the former further decreases to $3\%$ and $2\%$ for the weighted and unweighted delay, in average. All improvements are statistically significant according to the Wilcoxon signed-rank test with $\alpha=0.05$. The comparison between SO and CEN* shows that the former, in the short computational time it is allowed to use, typically gets extremely close to the optimal solution, or to the best solution found by the state-of-the-art RECIFE-MILP in a one-hour execution. Indeed, this time is not always enough to find the optimal solution, as some traffic situations result in extremely difficult instances: in $81\%$ out of the total 2150 executions, CEN* fails to prove solution optimality, and gaps remain between $0.1\%$ and $13.2\%$ ($0.5\%$ in average). In all scenarios, there is at least one execution in which optimality is not proven. In $11$ scenarios, SO manages to deal with the difficult instances more effectively, improving over CEN* in both weighted and unweighted delay. According to the Wilcoxon signed-rank test with $\alpha=0.05$, the difference remains statistically significant in favor of CEN* when unweighted delay is considered. Instead, no significant difference is observed in terms of weighted delay. 

The very large improvements observed with respect to FCFS show that the scenarios are actually quite complex, and optimization allows a much more effective traffic management than the application of the myopic rule.
However, all  optimization-based approaches do not achieve the same traffic effectiveness, SO being  extremely close to CEN* and almost always slightly better than  both CEN-1 and CEN-2 and reaching an improvement of around $20\%$ in the best case, regardless of the type of delay considered. 

There are multiple factors that could explain how this improvement can be achieved.
First, we analyze the impact of the different objective functions used for SO and  CENs, that is, the unweighted versus weighted delay. Indeed,  CENs consider  the former, while SO considers the latter in each train's hypothesis generation, even if the weights of the trains different from the focal one are not precisely known. Although the merge repair operation described in Section~\ref{subsec:merge} is performed centrally in SO, and as such minimizes unweighted delays in the objective function, we consider the impact of this operation minimal, also knowing that all repair actions are re-assessed by trains at later stages, thus based on weighted delays.
For this analysis, we study the conflict resolutions decided by both SO and CEN-1  at each execution  and for each scenario. The explicit information of what conflicts are dealt with when making decisions is not available, as the solution generation procedure employed is based on branch and bound and does not specifically identify conflicts. Hence, we define a proxy for the set of conflict resolutions decided, which we refer to as \textit{quasi-conflicts}. At each execution, given the RTTP produced by the considered TMS (either CEN-1  or SO), we identify all pairs of trains in which one or both exit the control area with a positive delay, and which have paths that immediately follow each other on at least one TDS. If the paths of a pair of trains follow each other immediately on more than a TDS, we consider the first one. These situations correspond to the quasi-conflicts we use as a proxy for the set of conflicts solved by the TMS. In the $35$ scenarios, CEN-1  returns RTTPs with $517$ quasi-conflicts  in average (between $359$ and $747$), while SO with $424$ (between $336$ and $526$). The pairs of trains involved in these quasi-conflicts are at most $140$ for CEN-1  and $132$ for SO, with averages across scenarios of $121$ and $110$. The higher quantities observed for CEN-1  are consistent with the worse performance observed for this approach in terms of total delay, be it weighted or unweighted: although there is no strictly monotonic relation, often more conflicts (and therefore quasi-conflicts) result in higher delays. When we look more closely at these quasi-conflicts, we see that, in average, both CEN-1  and SO make the train with the higher weight pass first in $46\%$ of these cases. 
The train with the lower weight passes first in $48\%$ of the cases for CEN-1  and in $49\%$ of the cases for SO, in average, the remaining quasi-conflicts being between trains with equal weight. 
From these numbers, it emerges that SO does not appear to favor higher-weight trains. Indeed, in these scenarios, the use of different weights for train delay in the objective function does not seem to impact SO choices  to the point of having it inverting train passing orders in many conflicts. Overall, we can conclude that the improvement of SO over CEN-1  does not depend on the different objective functions used. 

Another relevant factor that differentiates SO and CENs  is problem decomposition. Indeed, in the former, trains generate hypotheses that possibly modify only the paths of their neighbors. They hence operate on sub-instances of the rtRTMP, while CENs  considers all paths as modifiable and operates on complete instances.
On the one hand, this intuitively makes it easier for SO to find the optimal solution for each traffic management sub-instance trains happen to encounter. But, on the other hand, the choices made could be myopic, as they do not consider possible conflicts in the long run. In particular, in SO, trains may make decisions that are good regarding the first trains they may interact with, but that may result not to be good later on. 

To better understand how problem decomposition affects the overall solutions, it is useful to look at how the traffic evolves over time in the proposed scenarios, and how this impacts the quality of the solutions produced by RECIFE-MILP and its hypothesis generation variant. 
\begin{figure}[!b]
    \centering
    \includegraphics[width=\columnwidth]{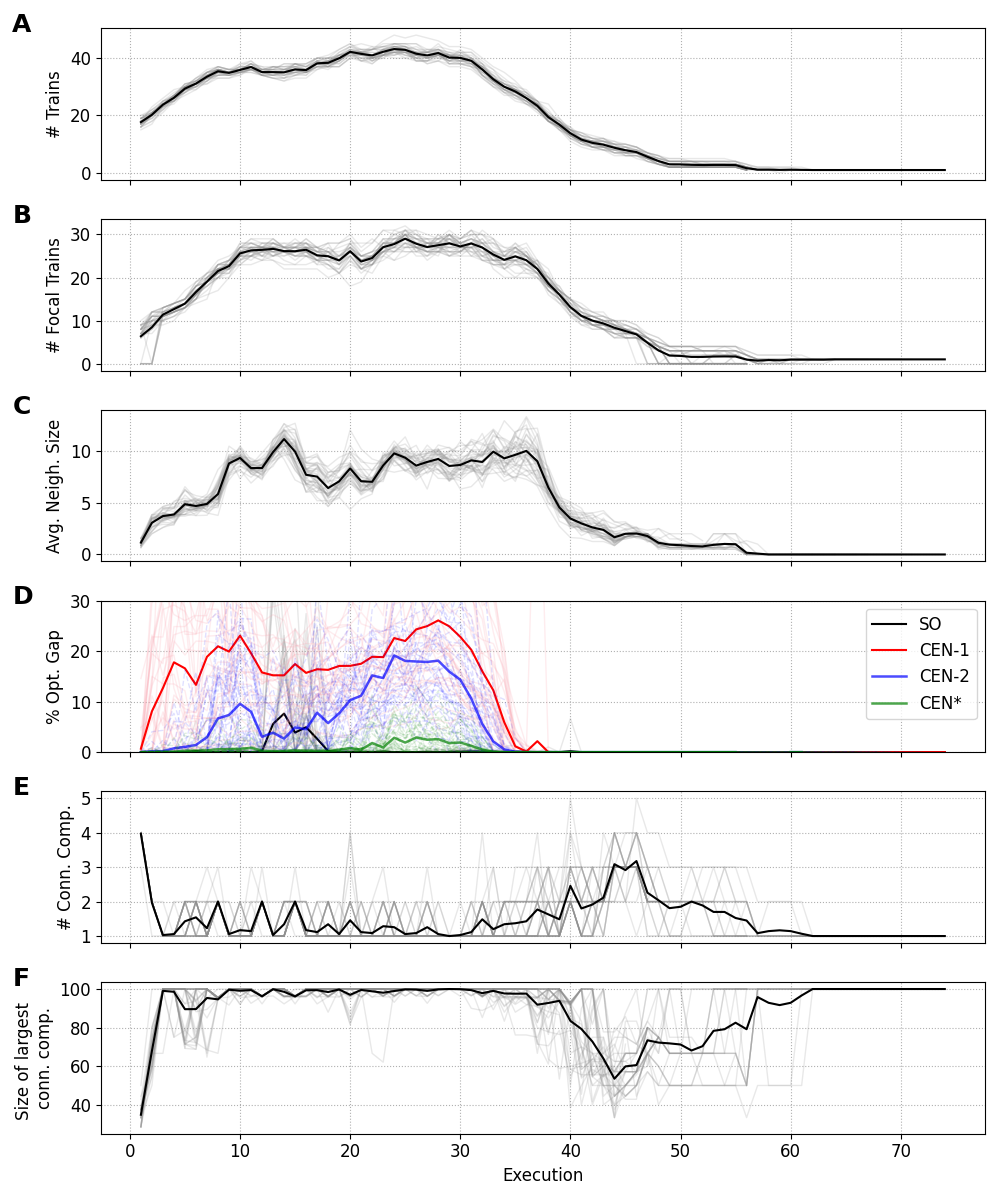}
    \caption{Representation of traffic and its management evolution across scenarios and executions: thick lines represent average values across scenarios while thin lines represent individual scenarios. \textbf{A}: Total number of trains. \textbf{B}: Number of focal trains. \textbf{C}: Average size of the neighborhood of a train. \textbf{D}: Percentage optimality gap per execution, for RECIFE-MILP in CEN-1 and 2, and CEN*  and its hypothesis generation variant in SO. \textbf{E}: Number of connected components. \textbf{F}: percentage of focal trains included in the largest connected component.}
    \label{fig:info_by_iter}
\end{figure}
Figure~\ref{fig:info_by_iter} shows how the traffic and its management evolve over time. During the simulated period, the number of trains considered in the traffic management (i.e., those that are in the control area or will enter it in a time window of 40 minutes) reaches a maximum value of about $40$ trains (see Figure~\ref{fig:info_by_iter}A). Of these trains, in average, at most $\sim 25$ are focal: they are present in the control area or will enter it within the neighborhood selection time horizon ($T_h=15$ minutes) and, as such, they generate hypotheses and actively participate to the traffic management process in SO (see Figure~\ref{fig:info_by_iter}B). For each focal train, the neighborhood size is rather small, with $\sim 10$ trains in average, suggesting that problem decomposition can actually reduce the size of the instances (see Figure~\ref{fig:info_by_iter}C).
After execution  35 (around \mbox{4:00}~AM), these numbers start to decrease as this is the end of the considered horizon, no more trains enter the control area and the ones that are present progressively exit it. Executions  between 0 and 35 are referred to as the \emph{regime interval}, while executions  beyond this interval are referred to as the \emph{tail} of the simulated scenario. 
%
%
%
%
The fact that train neighborhoods are rather small, even in the regime interval, leads to a higher quality of the produced solutions. Figure~\ref{fig:info_by_iter}D reports the \textit{percentage optimality gap} achieved by RECIFE-MILP in CEN-1, 2 and CEN*,  and by its hypothesis generation variant in SO along executions: this is the difference between the value of the best solution found and the lower bound computed during the branch and bound procedure. 
Clearly, SO benefits from problem decomposition, achieving optimality in almost all cases, while CENs are  often penalized by the difficulty of the overall traffic situation and typically returns sub-optimal solutions with gaps of almost $20\%$ in average for CEN-1, and $4\%$ for CEN-2 that operates on a smaller search space thanks to the initial reduction of the number of alternative routes through ACO-TRSP. CEN*, despite its much longer computation time, registers an average gap of $0.5\%$.

The impact of myopia due to problem decomposition is very difficult to quantify, as the fact of making decisions periodically in closed-loop partially mitigates it. As a proxy for the impact of myopic decisions at each execution, though, we look at the infeasibilities that arise after the merge operations, which require an intervention to repair the RTTP. 
Our analysis reveals that such repair procedures is required in only $3.5\%$ of the cases. This indicates that SO effectively manages traffic even in the long run, thanks to the consideration of the paths of trains out of the neighborhoods as hard constraints  in the hypothesis generation module. This reduces the need for centralized intervention only to a handful of cases, in which trains not identified as potentially in conflict in the neighborhood selection module change their path in ways that create inconsistencies in the far-away future (see Section~\ref{subsec:merge}). 

Another element possibly impacting the myopia of decisions is the number and size of the connected components of $G_I$ (see Figure~\ref{fig:info_by_iter}E and ~\ref{fig:info_by_iter}F). Indeed, if all trains were part of the same connected component of the graph, their decisions would be all consistent, at least to some extent. In our experiments, we observe that in the regime interval there often is only one large connected component, involving almost all the trains present in the control area or entering it within $T_h$. When more than one connected component exists and the maximum size of these components is around $100\%$, it means that one of the components is actually made of one or two trains, while the other includes all the remaining focal trains. The number of connected components increases up to five only in the tail, when trains  start to exit the control area. We then consider, for all perturbation scenarios, the relationship between the number of connected components and the occurrence of repair procedures, limited to the executions  in which the repair procedure was necessary. In other words, given all the executions  in which at least one perturbation scenario resulted in a final repair procedure, we compute the number of connected components for all perturbation scenarios, distinguishing between those where repair was necessary and those where it was not. 
Here, a Mann-Whitney U test does not support a statistically significant difference between the number of connected components in instances requiring merge repair and those that do not ($p\text{-value} = 0.77$). This suggests that the 
graph connectivity is not the primary driver for merge repair interventions, which instead could depend on the degree of nodes in the interaction graph combined with the designed hypothesis compatibility, that limits the detection of conflicts only for pairs involving the focal trains. How different graph characteristics are linked to merge repair needs will be further investigated in future work.

In summary, problem decomposition appears to be a definite strength for SO with respect to CEN's, as no remarkable effect of myopic decision making is observed.

\begin{figure}
    \centering
    \includegraphics[width=\linewidth]{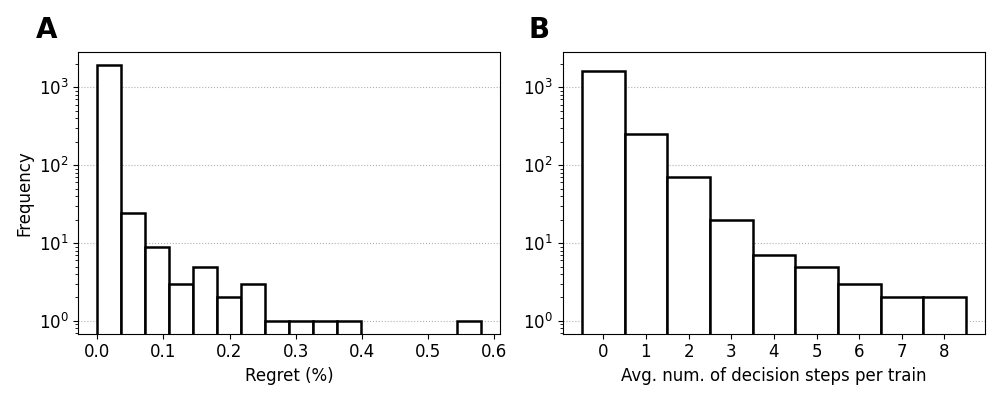}
    \caption{Analysis of the consensus process: \textbf{A}: Regret distribution. \textbf{B}: Distribution of the number of decision steps per train.}
    \label{fig:consensus-distr}
\end{figure}

To conclude the analysis of the results, we observe the performance of the consensus process designed for hypothesis selection. First of all, trains reach consensus on all the $C_T=1974$ calls of the hypothesis selection module ($100\%$). Figure~\ref{fig:consensus-distr}A shows the distribution of regret, i.e., the distribution of the difference between the total delay of the selected hypotheses with respect to the optimal one.
It emerges that in $C_O=1783$ cases ($90.3\%$), the consensus converges to the optimal solution, while, in the remaining cases, the maximum regret is only $0.6\%$. This signifies that even when the consensus solution is not optimal, it is very close to the optimal one.
The average number of decisions per train required to reach consensus follows the distribution shown in Figure \ref{fig:consensus-distr}B. 
In $C_I=1614$ cases ($81.8\%$), consensus is reached immediately, requiring no iterative adjustments. Given that trains initiate the hypothesis selection process by selecting their best hypothesis, an immediate consensus signifies that no hypothesis change was necessary: all selected hypotheses are compatible. Hence, the problem decomposition directly leads to a feasible solution. When further decision steps are necessary, the trains converge to a solution in at most $8$ steps, in average. Out of the $C_T-C_I=360$ cases in which at least one iteration of the hypothesis selection is required, $C_O-C_I=169$ cases achieve the optimal solution (i.e., null regret). Hence, nearly half of the times ($\frac{C_O-C_I}{C_T-C_I}=47\%$), executing the consensus process achieves optimality.

For an analysis of the performance of the consensus process in abstract complex networks, see~\cite{d2025decentralised}.



\section{Conclusions and future work}\label{sec:conclusion}

In this paper, we proposed an innovative approach to railway traffic management. Specifically, we designed a comprehensive process for having trains self-organizing to minimize delays in case of perturbations. To the best of our knowledge, this is the first proposal of such a process that consistently produces feasible and optimized traffic management plans and allows a realistic assessment. 
For this assessment, we considered a closed-loop framework in which a microscopic simulator replaces reality and all decisions are made through the self-organizing process.

In a thorough experimental analysis covering a portion of the Italian network shared by different railway undertakings operating mixed traffic, we showed that the traffic self-organization  often found solutions that are very close to the optimal ones,  and   was able to outperform the state-of-the-art centralized decision-making algorithm used as a benchmark. This result was achieved thanks to the careful design of the various modules of the self-organization process. These modules allow the suitable instance decomposition, to profit from the increased problem tractability without paying the price of making myopic decisions.

In future research, we will deeply investigate the impact of each module configuration on the self-organization performance, also linked to different types and sources of traffic perturbations. Moreover, we will propose alternative designs for some of them, relying for example on machine learning techniques to further improve the instance decomposition,
 
and we will exploit graph theoretical analysis to try to explain the relations between modules' configuration and design, and system performance. 
The impact of different traffic characteristics also deserves future studies, in particular to understand the effect of having a few competing companies, each operating many trains. 
%

\section{Acknowledgments}
\noindent The work presented in this paper has been carried out in the context of the SORTEDMOBILITY project. This project is supported by the European Commission and
funded under the Horizon 2020 ERA-NET Cofund scheme
under grant agreement N 875022. Vito Trianni and Leo D'Amato acknowledge partial support by TAILOR, a project funded by EU Horizon 2020 research and innovation program under GA No 952215.

\bibliographystyle{IEEEtran}
\bibliography{References}

\begin{IEEEbiography}[{\includegraphics[width=1in,height=1.25in,clip,keepaspectratio]{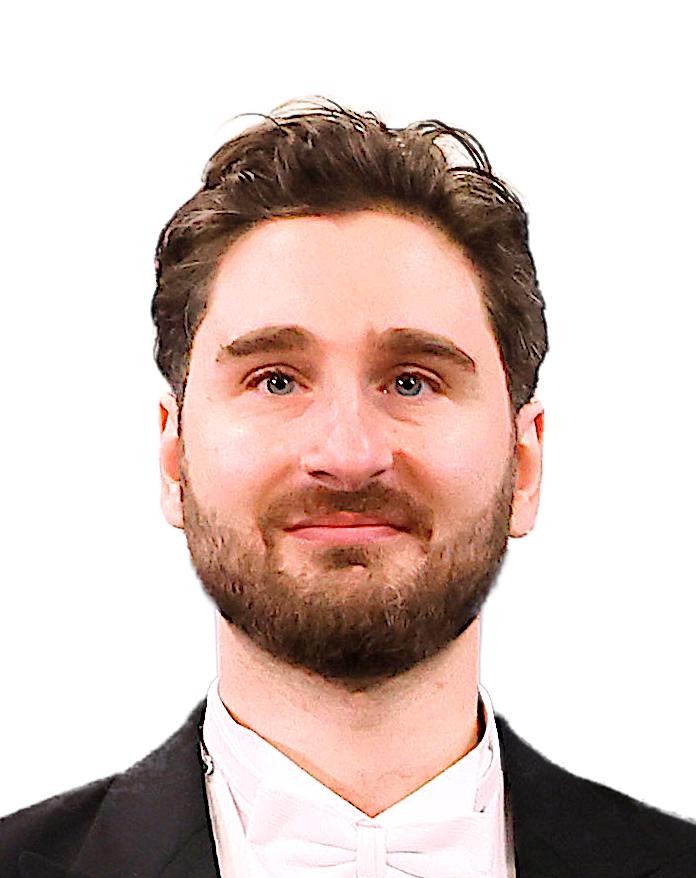}}]{Federico Naldini} \hyphenpenalty=9000 \exhyphenpenalty=9000
is a computer science researcher at ESTAS, Université Gustave Eiffel,
France. He received a PhD in Operations Research and an MSc in Industrial
Engineering from the University of Bologna, Italy. He combines software
engineering with optimization, AI, metaheuristics, and simulation to
develop decision-support systems for industrial and transportation
applications. His research focuses on real-time rail traffic management,
including train rescheduling, rerouting, and energy-efficient operations.
\end{IEEEbiography}

\begin{IEEEbiography}[{\includegraphics[width=1in,height=1.25in,clip,keepaspectratio]{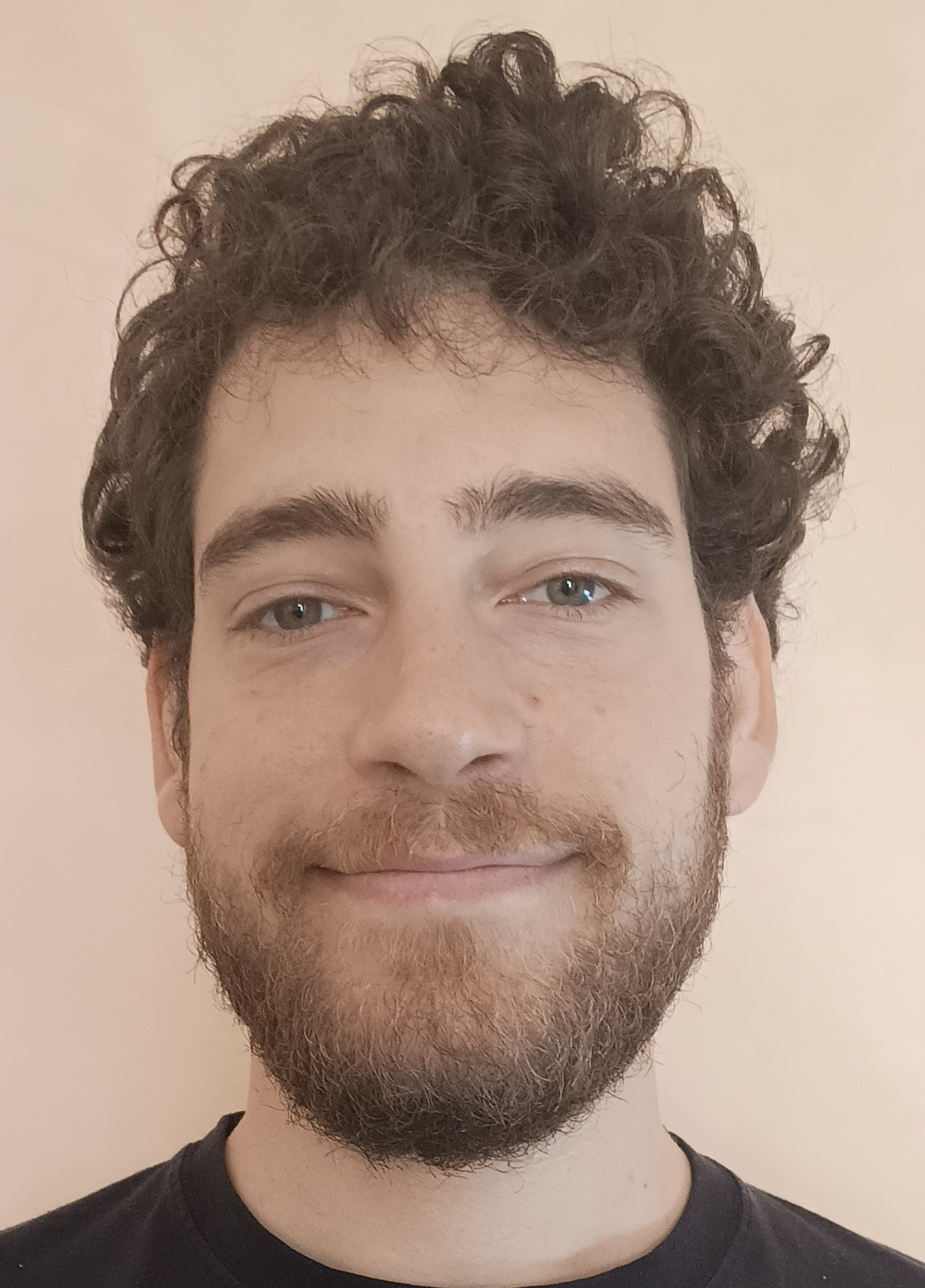}}]{Fabio Oddi}
earned a Master's degree in Control Engineering from Sapienza University of Rome, Italy, in 2022. He is currently pursuing a Ph.D. in Computer Science Engineering at the same institution, with his research funded by the Institute of Cognitive Sciences and Technologies of the National Research Council (ISTC-CNR). His research focuses on exploring the role of interaction networks in decision-making processes for decentralized agent-based systems.\end{IEEEbiography}

\begin{IEEEbiography}[{\includegraphics[width=1in,height=1.25in,clip,keepaspectratio]{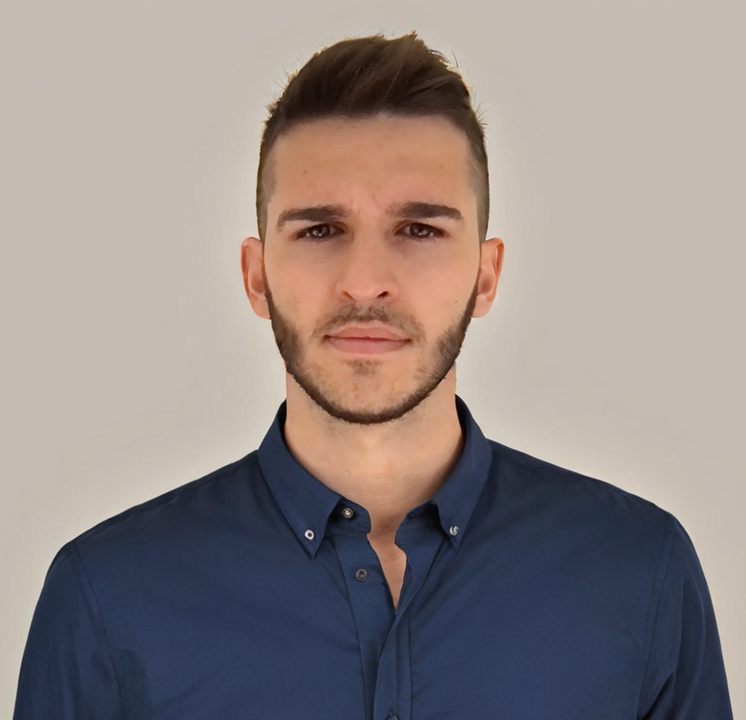}}]{Leo D'Amato}
is a PhD candidate in Artificial Intelligence at the Polytechnic University of Turin and a research fellow at the Institute of Cognitive Sciences and Technologies (ISTC-CNR) of the National Research Council of Italy. His research mainly focuses on AI, multi-agent systems and cognitive science. \end{IEEEbiography}

\begin{IEEEbiography}[{\includegraphics[width=1in,height=1.25in,clip,keepaspectratio]{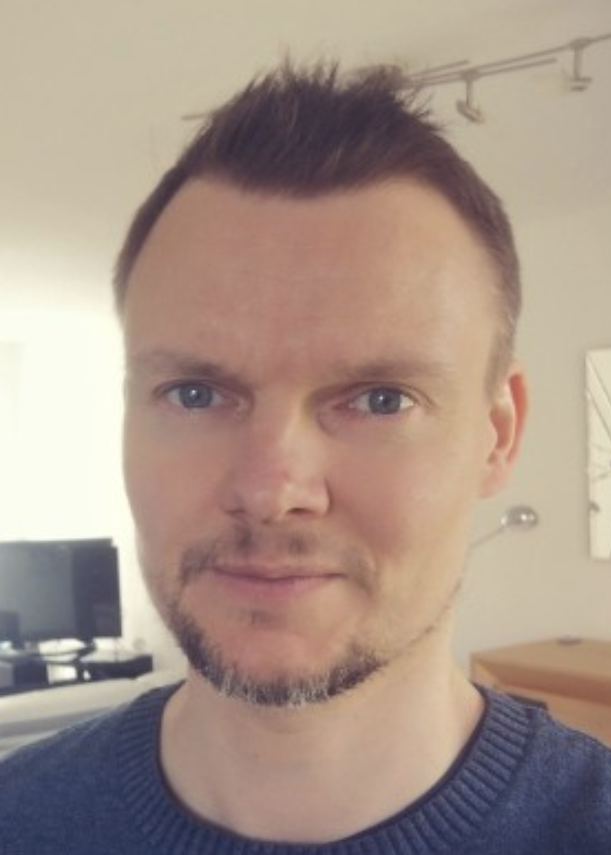}}]{Grégory Marlière} is an engineer at Université Gustave Eiffel, France, specializing in optimization and computer science for railway operations. His expertise focuses on developing decision support systems for real-time railway traffic management.
He is co-designer of several mathematical models including RECIFE-MILP and RECIFE-CPI, which apply mixed-integer linear programming and constraint programming techniques to railway optimization problems. Grégory leads the development of the RECIFE simulation platform that connects optimization models with railway infrastructure databases and traffic simulators.
\end{IEEEbiography}

\begin{IEEEbiography}[{\includegraphics[width=1in,height=1.25in,clip,keepaspectratio]{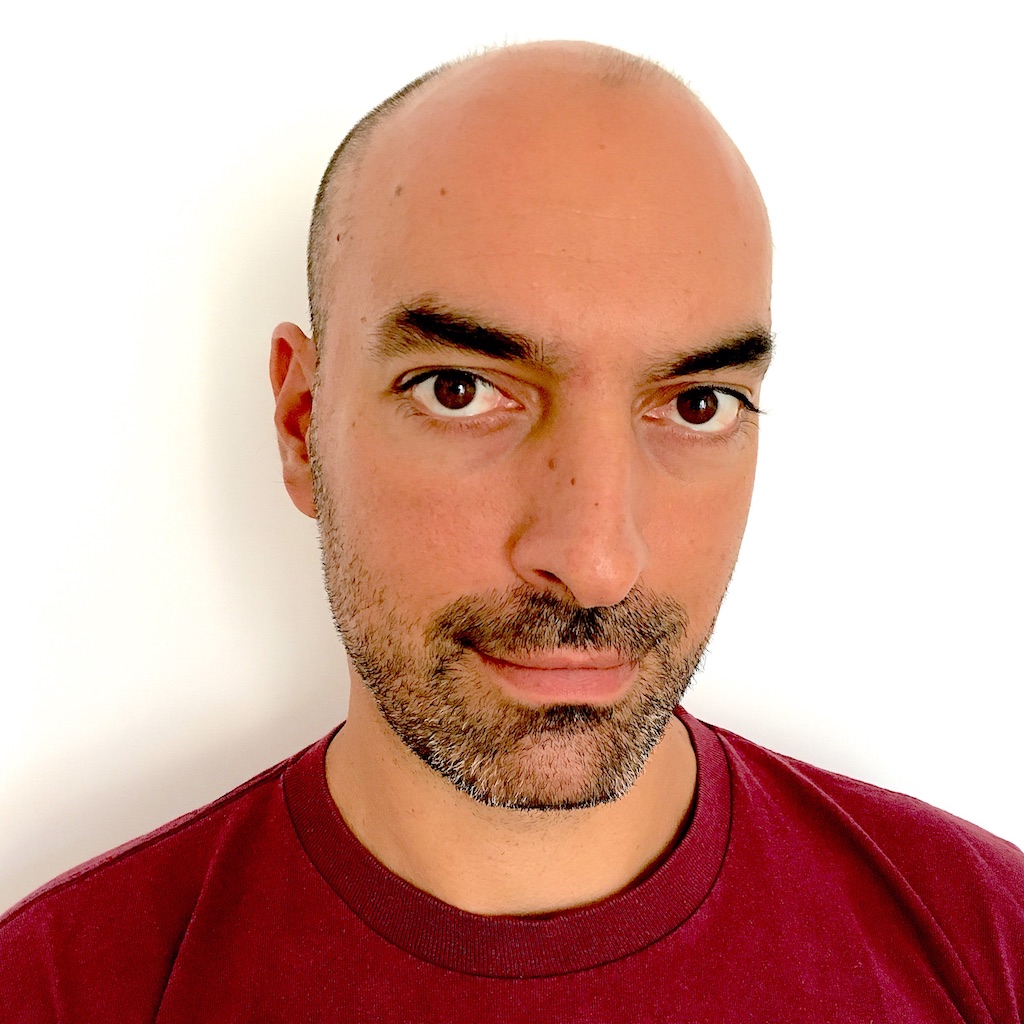}}]{Vito Trianni} is a researcher director at the Institute of Cognitive Sciences and Technologies of the Italian National Research Council (ISTC-CNR). He received the Ph.D. in Applied Sciences at the Université Libre de Bruxelles (Belgium) in 2006. 
Vito Trianni’s research mainly involves AI and robotics, with particular emphasis on collective intelligence and self-organising systems. He conducts research on the analysis and design of large-scale decentralized systems, including robot swarms, cyber-physical and socio-technical systems. Finally, he is interested in the study of hybrid collective intelligence, where humans and machines collaborate in groups for decision support in high-stake domains, such as medical diagnostics or policy making for climate change adaptation management.\end{IEEEbiography}

\begin{IEEEbiography}[{\includegraphics[width=1in,height=1.25in,clip,keepaspectratio]{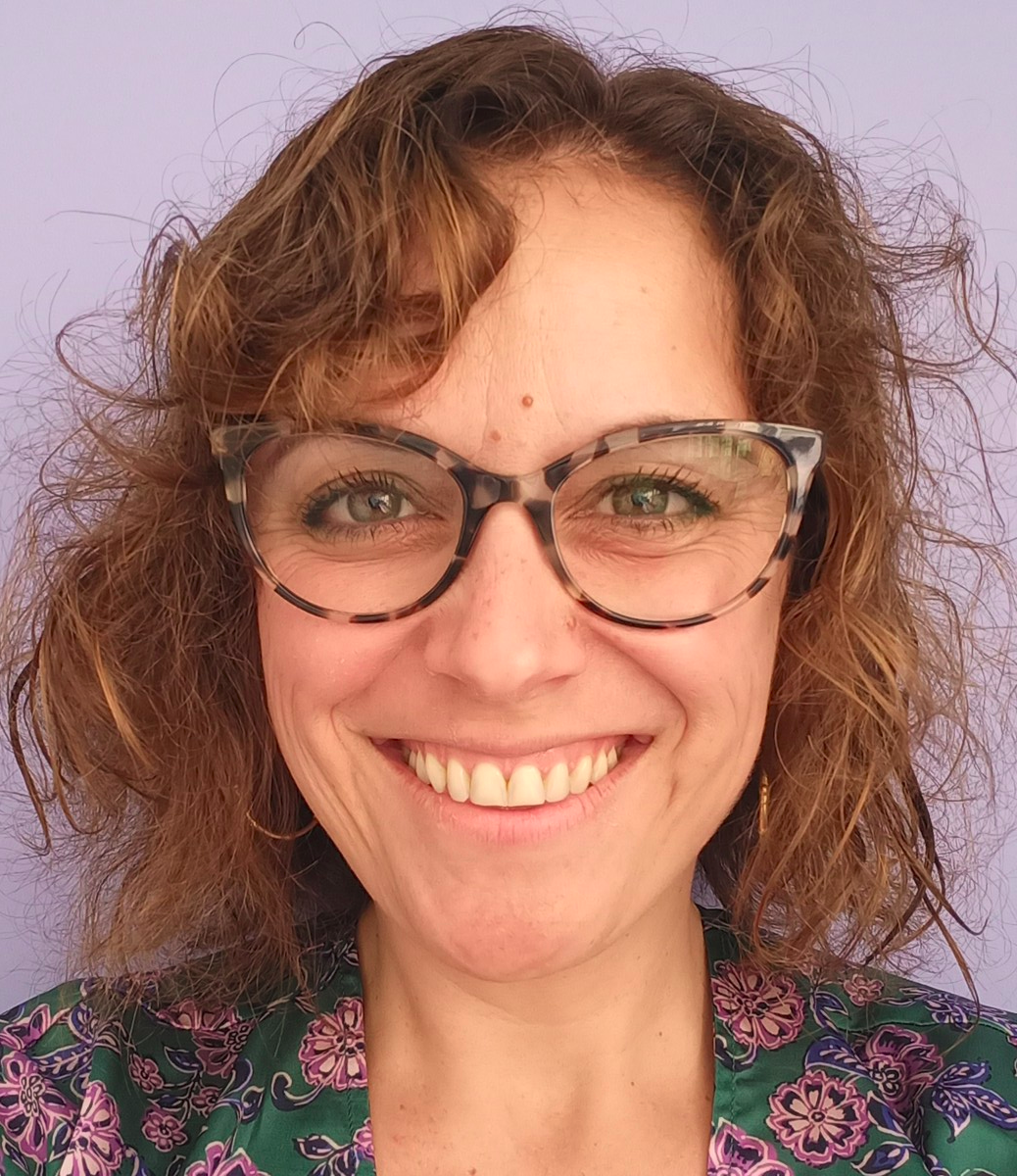}}]{Paola Pellegrini}
is an expert in railway planning and operational management. She received her PhD in 2007 at Ca' Foscari University in Venice, Italy. Her research focuses on the development of optimization approaches to effectively exploit railway infrastructure capacity, aiming to the automation of the involved processes. Paola has participated in a number of national and international research projects, and she regularly collaborates with various SMEs to facilitate the knowledge transfers from academia to practice. She is a member of the Board of the International Association of Railway Operations Research (IAROR).\end{IEEEbiography}
\end{document}